# Robust Joint Active-Passive Beamforming Design for IRS-Assisted ISAC Systems

Mahmoud AlaaEldin, *Student Member, IEEE*, Emad Alsusa, *Senior Member, IEEE*, Karim G. Seddik, *Senior Member, IEEE*, Christos Masouros, *Senior Member, IEEE*, and Iman Valiulahi, *Student Member, IEEE*

*Abstract*—The idea of Integrated Sensing and Communication (ISAC) offers a promising solution to the problem of spectrum congestion in future wireless networks. This paper studies the integration of intelligent reflective surfaces (IRS) with ISAC systems to improve the performance of radar and communication services. Specifically, an IRS-assisted ISAC system is investigated where a multi-antenna base station (BS) performs multi-target detection and multi-user communication. A low complexity and efficient joint optimization of transmit beamforming at the BS and reflective beamforming at the IRS is proposed. This is done by jointly optimizing the BS beamformers and IRS reflection coefficients to minimize the Frobenius distance between the covariance matrices of the transmitted signal and the desired radar beam pattern. This optimization aims to satisfy the signal-to-interference-and-noise ratio (SINR) constraints of the communication users, the total transmit power limit at the BS, and the unit modulus constraints of the IRS reflection coefficients. To address the resulting complex non-convex optimization problem, an efficient alternating optimization (AO) algorithm combining fractional programming (FP), semi-definite programming (SDP), and second order cone programming (SOCP) methods is proposed. Furthermore, we propose robust beamforming optimization for IRS-ISAC systems by adapting the proposed optimization algorithm to the IRS channel uncertainties that may exist in practical systems. Using advanced tools from convex optimization theory, the constraints containing uncertainty are transformed to their equivalent linear matrix inequalities (LMIs) to account for the channels' uncertainty radius. The results presented quantify the benefits of IRS-ISAC systems under various conditions and demonstrate the effectiveness of the proposed algorithm.

## I. INTRODUCTION

The advent of integrated sensing and communications (ISAC) has been recognized as one of the key technologies for beyond-fifth-generation (B5G) and sixth-generation (6G) wireless networks to enable environment-aware applications such as auto-driving, industrial automation, and mixed reality, in which the wireless infrastructures and spectrum resources are reused for radar sensing, localization, and imaging [1]–[3]. To enable sensing and communication services, various ISAC system architectures, waveforms, and transmit beamforming designs have been proposed in prior works, [4]–[11].

Similarly, intelligent reflective surfaces (IRS) has recently emerged as a promising technology within the wireless communication community [12]–[15]. IRS comprises a planar surface equipped with numerous adjustable and low-cost passive reflecting elements, capable of modifying amplitudes and phase shifts to reflect incident signals in a controlled manner. This technology is hugely appealing due to its potential to significantly enhance the spectral and energy efficiency of wireless networks [16]. The fundamental merit of IRS lies in its ability to manipulate the wireless channel between the transmitter and receiver to satisfy diverse design objectives, including signal enhancement and interference reduction.

IRS has been proposed in many wireless communication systems in recent years as a means to boost the performance and efficiency of such systems. For instance, in [17], IRS was proposed to enhance the performance of a single cell communication system involving a multi-antenna base station (BS) which serves multiple single-antenna users. In another study [18], the authors proposed efficient algorithms to maximize the energy efficiency (EE) of IRS-assisted downlink communication systems, involving a multi-antenna BS that serves multiple users, by jointly optimizing the transmit power allocation at BS and the IRS phase shifts. IRS has also found applications in simultaneous wireless information and power transfer (SWIPT) systems [19]. In [20], IRS was employed to improve the energy harvesting performance of SWIPT systems, where a multi-antenna BS simultaneously transmitted information and energy to several receivers. The use of IRS in an uplink multi-user scenario to optimize spatial multiplexing with a linear receiver was explored in [21]. Additionally, the authors of [22] proposed a joint power and user association scheme for a multi-IRS enabled multi-BS downlink millimeter-wave system which communicates with multiple users. The versatility of IRS has led to applications in various domains, such as physical layer network coding [23], wireless mesh back-hauling [24], unmanned aerial vehicles (UAV) communications [25], physical layer security for covert communications [26], and enhancing the performance of ISAC systems as presented in details in the next subsection. As such, IRS is considered a pivotal enabling technology for the sixth generation (6G) wireless networks [27].

Mahmoud AlaaEldin and Emad Alsusa are with the Electrical and Electronic Engineering Department, University of Manchester, Manchester, M13 9PL, UK (e-mail: mahmoud.alaaeldin@manchester.ac.uk; e.alsusa@manchester.ac.uk).

Karim G. Seddik is with the Department of Electronics and Communications Engineering, American University in Cairo, Cairo, Egypt 11835 (e-mail: kseddik@aucegypt.edu).

C. Masouros and Iman Valiulahi are with the Department of Electronic and Electrical Engineering, University College London, London, WC1E 7JE, UK (e-mail: i.valiulahi@ucl.ac.uk; c.masouros@ucl.ac.uk).

*A. Literature review on IRS-assisted ISAC systems*

Numerous spectrum-sharing methods have been proposed in the literature to address the issue of minimizing the mutual interference between the radar and communication systems. The main challenge lies in preventing interference when both radar and communication systems are operating at the same time to ensure satisfactory quality-of-service (QoS). Therefore, several works in the literature studied the usage of IRS to address the mutual interference problem in ISAC and enhance the performance of both radar and communications services. These works proposed different solutions to jointly solve the transmit and reflective beamforming problems at BS and IRS.

The authors in [28] considered a system where a multiple input multiple output (MIMO) radar coexists with a multi-user multiple input single output (MISO) communication BS. They exploited IRS to mitigate the interference coming from the communication BS to the radar system by maximizing the radar detection probability subject to the users' signal-to-interference-plus-noise-ratio (SINR) constraints. In [29], a joint waveform design and passive beamforming in IRS-assisted ISAC system was investigated to minimize the multi-user interference (MUI) of the communication users under a strict beampattern constraint on the transmitted radar-communication signal. However, the authors did not consider individual SINR constraints for the communication users to simplify the problem which may raise fairness issues. Further, the ISAC waveform design was studied in [30] by minimizing the MUI under Cramer-Rao bound constraint for the radar side. Furthermore, the authors in [31] extended the work in [29], [30] by maximizing the SINR for the radar and meanwhile minimizing the MUI of the communication users. They exploited block coordinate descent and Element-wise Closed-Form (ECF) techniques to optimize the waveform and the IRS reflection coefficients. Moreover, the authors in [32] considered maximizing the communication data rate and the mutual information for the radar sensing by jointly optimizing the transmit beamforming at BS and the reflection coefficients at IRS. Additionally, a double-IRS-assisted radar-communication coexistence system was proposed in [33] where two IRS panels were deployed to enhance the communication signals and mitigate the mutual interference between the radar and communication systems. The authors in [34] jointly optimized active beamforming at BS and passive beamforming at IRS to maximize the sum-rate of the communication users while satisfying a beampattern similarity constraint for the radar. However, in [35], maximization of the weighted sum of target detection SINRs was considered to optimize the IRS-assisted ISAC system, while satisfying the SINR constraints of the communication users. It is noteworthy that manifold optimization based algorithms were utilized in [34], [35] to optimize the IRS reflection coefficients. The work in [35] was extended in [36] to consider an active IRS-aided ISAC system where active IRS can help with the multiplicative fading effect when the target-BS direct link is not available. Moreover, the authors in [37] considered the sum-rate maximization of the communication system that coexists with a separate radar BS. They considered joint optimization of the beamformer at BS and IRS reflection coefficients, subject to maximum allowable interference power from the communication BS to the radar system. In [38], a hybrid IRS comprising of active and passive elements was used to assist an ISAC system, where the authors jointly optimized transmit beamformers and IRS coefficients, i.e., active and passive elements, to maximize the worst-case target illumination power. In [39], a secure IRS-aided ISAC system was investigated where the target may be a suspicious eavesdropper that probably spies out information that concerns the communication users. The authors used dedicated sensing signals that are transmitted by BS other than the communication signals to ensure the sensing quality.

Although all above reviewed works have studied IRS-assisted ISAC system, non of those works investigated robust beamforming designs for such systems. The robust beamforming design is of high importance since the assumption of having perfect channel knowledge is impractical and can not be realized. Therefore, there will be always uncertainty (error) in the IRS channel vectors that must be accounted for when performing the joint beamforming optimization at BS and IRS. It is also worthy investigating the impact of channel uncertainty on the performance of the IRS-assisted ISAC system and how such ambiguity can affect both subsystems. Thus, it is important to develop specific optimization techniques for the beamforming of the IRS-assisted ISAC systems that can be adjusted to deal with such channel uncertainty without violating the constraints of the optimization problem.

*B. Motivations and Contributions*

Motivated by the potential of ISAC, in this paper, we investigate robust joint beamforming design for the IRS assisted ISAC system considering the shared deployment, where all the BS antennas are used for both the radar and communications services. In the shared deployment scheme, the BS transmits only one signal that has the desired beam pattern similarity for radar sensing and also carries data to the communication users in the system. The shared deployment scheme of the IRS-assisted ISAC system is worth studying since it reduces the installation cost by canceling the deployment of a separate BS for the radar service and using the main BS for both radar and communications services. Therefore, the ISAC shared deployment scheme has the potential to be an integrated part of future wireless networks which will provide both communications and localization services simultaneously.

The transmitted ISAC signal is the summation of the active beamformers at the BS antenna array. Thus, these active beamformers have to be designed to have the desired beam pattern characteristics of the radar sensing, and at the same time, to carry information to the communication users with the desired QoS requirements. To achieve this, active beamforming at BS and reflective beamforming at IRS are jointly optimized so that the Frobenius distance between the covariance matrices of the transmitted signal and the desired radar beampattern is minimized while satisfying the required SINR constraints of the communication users. The fundamental trade off of our problem is that the BS splits its total power during the beamforming optimization between satisfying the communication data rates and approaching the desired radar beampattern covariance matrix. Consequently, the usage of IRS in this problem is very beneficial by assisting the communication users to achieve the desired data rates, which in turn allows the BS to put more of its power budget in enhancing the radar sensing performance. The resulting optimization problem is non-convex and challenging to solve due to the coupling of the optimization variables and having non-convex cost function. Therefore, we develop an efficient and low complexity alternation optimization (AO) algorithm based on fractional programming (FP), semi-definite relaxation (SDR), and second order cone programming (SOCP) optimization methods to transform the resulting non-convex optimization problem into two convex subproblems that can be solved using CVX in an iterative manner using AO. The developed optimization algorithm is specifically designed using convex programming so it can be adjusted to account for the IRS channel uncertainty in practical systems. This is unlike similar works in [34], [35] which considered the IRS-ISAC shared deployment too by performing the beamforming optimization using manifold optimization techniques. However, manifold optimization techniques cannot be adjusted to consider the channel uncertainties for robust beamforming design which distinguish our work in this manuscript. Moreover, complexity analysis for the proposed optimization algorithm is presented and compared against the SDR benchmark. Thus, the main contributions of this work can be listed as follows

- An efficient and low complexity optimization algorithm is proposed for IRS-enabled ISAC systems to jointly optimize the transmit beamforming at BS and reflective beamforming at IRS. The optimization aims at minimizing the difference between the covariance matrices of the transmitted radar-communication signal and the desired radar beampattern, while maintaining the SINR constraints of the communication users.
- To solve the resulting non-convex beamforming problem, an AO is devised to solve the active beamforming at BS and reflective beamforming at IRS in an iterative manner. The proposed algorithm combines FP, semi-definite programming (SDP) and SOCP to convert the non-convex cost function and constraints of the problem into convex counterparts.
- The proposed convex optimization based algorithm is further extended to consider the robust beamforming problem for IRS-ISAC systems where no perfect channel knowledge can be available at BS. This is done by converting the constraints having uncertainty into their equivalent linear matrix inequality (LMI)s which transform the feasible set of the problem into an intersection of convex cones. The proposed robust technique can ensure satisfying the constraints of the problem in the presence of channel uncertainties.
- Finally, simulation results are provided to demonstrate the effectiveness of the proposed algorithm for both perfect and uncertain channel knowledge cases. The mean squared error (MSE) and peak to side-lope ratio (PSLR) of the obtained beampatterns using the proposed algorithm outperforms those obtained by SDR based benchmark, with less computational complexity. Furthermore, we examine the performance loss due to having different channel uncertainty radii, and see their effect on the obtained beampatterns.

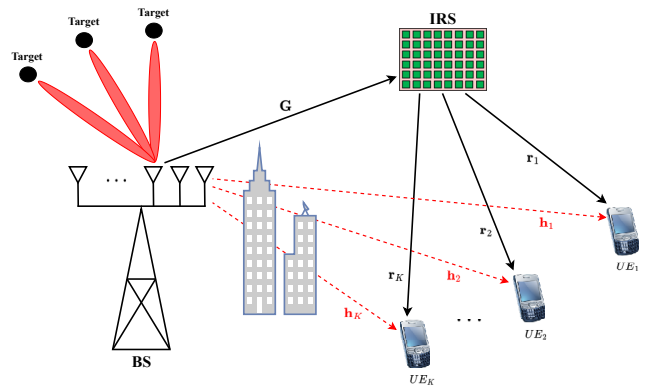

Figure 1: IRS assisted ISAC system model

## C. Organization and Notations

The rest of the paper is organized as follows. In Sec. II, the IRS-aided ISAC system model is presented. Sec. III presents the beamforming problem formulation of the presented system model. The proposed solution of the problem for the perfect channel knowledge case is discussed in Sec. IV, while the robust beamforming technique is presented in Sec. V for the case of having channel uncertainty. Finally, the results and conclusions are given in Sections VI and VII, respectively.

*Notations*: Throughout this paper, scalars are denoted by lower-case letters. Bold lowercase letters and bold uppercase letters denote vectors and matrices, respectively. The notations $(.)^T$, $(.)^*$ and $(.)^H$ represent transpose, conjugate and hermitian of a matrix or a vector, respectively. The operators $\|.\|_F$ and $\|.\|_1$ represent the Frobenius norm and the $l_1$-norm, respectively.

## II. SYSTEM MODEL

As shown in Fig. 1, we consider an IRS-assisted ISAC system in which a multi-antenna BS, having $N$ antennas, simultaneously sends radar probing signals and transmits data to a group of $K$ single antenna users. The downlink data transmission is assisted by an IRS panel which consists of $L$ reflecting elements where the $i$th element can introduce a phase shift $\theta_i$ to the incident signal before reflection. The direct channels between the BS and the users are assumed to be independent and identically distributed (i.i.d.) Rayleigh fading channels as the line-of-sight (LoS) path may not exist, and the direct channel vector between user $k$ and the BS is denoted as $\mathbf{h}_k \in \mathbb{C}^{N \times 1}$. The BS-IRS and the IRS-user $k$ channels are denoted as $\mathbf{G} \in \mathbb{C}^{L \times N}$ and $\mathbf{r}_k \in \mathbb{C}^{L \times 1}$, respectively. The IRS panel is typically set up in a position where a LoS path to the BS and to the users can be exploited. Hence, we adopt Rician fading in this paper to model the channel vectors $\mathbf{g}$ and $\mathbf{h}_k$. The channel vector $\mathbf{g}$ is given as

$$\mathbf{G} = \sqrt{\frac{PL(d_{IB})K_{IB}}{K_{IB}+1}} \mathbf{G}^{LoS} + \sqrt{\frac{PL(d_{IB})}{K_{IB}+1}} \mathbf{G}^{NLoS}, \quad (1)$$

where $K_{IB}$ denotes the Rician factor of $\mathbf{g}$, $d_{IB}$ is the distance between the IRS and the BS, $\mathbf{G}^{LoS}$ and $\mathbf{G}^{NLoS}$ are the LoS and non-LoS (NLoS) components, respectively. The LoS component is deterministic, however, the elements of $\mathbf{g}^{NLoS}$ are i.i.d. complex normal, $\mathcal{CN}(0,1)$, random variables. The channel vector between user $k$ and the IRS, $\mathbf{r}_k$, is given as

$$\mathbf{r}_k = \sqrt{\frac{PL(d_{U_k,I})K_{UI}}{K_{UI}+1}}\mathbf{r}_k^{LoS} + \sqrt{\frac{PL(d_{U_k,I})}{K_{UI}+1}}\mathbf{r}_k^{NLoS}, \quad (2)$$

where $K_{UI}$ denotes the Rician factor of $\mathbf{r}_k$, $d_{U_k,I}$ is the distance between user $k$ and the IRS, $\mathbf{r}_k^{LoS}$ and $\mathbf{r}_k^{NLoS}$ are the LoS and NLoS components, respectively. The $PL$ factor in the channel model represents the path loss which is modeled for all the channels as

$$PL(d) = \eta_0 \left(\frac{d}{d_0}\right)^{-\alpha}, \quad (3)$$

where $\eta_0$ is the path loss at the reference distance $d_0 = 1\ m$, $d$ represents the link distance between the transmitter and the receiver, and $\alpha$ is the path loss exponent.

The BS precodes the data symbols of the $K$ users, then transmits the overall precoded signal to the users. All the $N$ antennas are shared for both radar detection and downlink communication as shown in Fig. 1. Therefore, the received signal at the $k$th user for this scenario is written as

$$y_k = \left(\mathbf{h}_k^T + \mathbf{r}_k^T \boldsymbol{\Theta} \mathbf{G}\right) \sum_{k=1}^{K} \mathbf{w}_k s_k + n_k, \quad (4)$$

where $s_k$ is the data symbol transmitted to user $k$, $\mathbb{E}[|s_k|^2] = 1$, $\mathbf{w}_k$ is the precoding vector of user $k$, and $n_k$ is the complex Gaussian $\mathcal{CN}(0, \sigma_n^2)$ additive white Gaussian noise (AWGN) at the $k$th user. The matrix, $\boldsymbol{\Theta}$, is an $L\times L$ diagonal matrix, i.e. $\boldsymbol{\Theta}=\text{diag}\{\mathbf{v}\}$, whose diagonal elements are the IRS reflection coefficients, where $\mathbf{v}=[e^{j\theta_1},\ldots,e^{j\theta_L}]^T$. The angle $\theta_i \in [0, 2\pi[$ represents the phase shift introduced by $i$th reflecting element.

The channels, $\mathbf{G}$, $\mathbf{r}_k$ and $\mathbf{h}_k$, are assumed to be known at BS with some uncertainty radius. The radar-communication system employs the communication signal as the radar probing waveform. The IRS panel here assists the communication users as it provides more coverage and strengthens the received signal at each user. This allows for designing the desired covariance matrix of the radar probing waveform while maintaining the QoS constraints for the communication users using the same transmit power at the BS. The covariance matrix of the transmitted radar-communication signal can be given as

$$\mathbf{C} = \sum_{k=1}^{K} \mathbf{W}_k, \quad (5)$$

where $\mathbf{W}_k = \mathbf{w}_k \mathbf{w}_k^H$. The total transmit power from the BS is given as

$$P = \sum_{k=1}^{K} \|\mathbf{w}_k\|^2 = \sum_{k=1}^{K} \text{Tr}\{\mathbf{W}_k\}, \quad (6)$$

where $\text{Tr}\{.\}$ denotes the trace of a matrix. The SINR at the $k$th user when it decodes its own signal can be expressed as

$$\gamma_k = \frac{\left|\left(\mathbf{h}_k^T + \mathbf{v}^T \text{diag}(\mathbf{r}_k)\mathbf{G}\right)\mathbf{w}_k\right|^2}{\sum_{\substack{i=1\\i\neq k}}^{K}\left|\left(\mathbf{h}_k^T + \mathbf{v}^T \text{diag}(\mathbf{r}_k)\mathbf{G}\right)\mathbf{w}_i\right|^2 + \sigma_n^2}, \quad (7)$$

where $\mathbf{v} \in \mathbb{C}^{L\times 1}$ is the passive beamforming vector that contains the IRS reflection coefficients. In the following section, the joint active-passive beamforming optimization problem of the IRS assisted ISAC system is presented.

### III. PROBLEM FORMULATION

In this section, we first briefly recall the radar beampattern design for the MIMO-radar only systems. Then, we present the IRS-assisted ISAC system problem formulation by combining the communication users' QoS constraints into the radar beampattern design problem. We also discuss how the IRS can enhance the performance of the radar service while maintaining the communication users' rate constraints.

*A. Radar only beampattern design*

The existing literature indicates that the design of such a beampattern is equivalent to designing the covariance matrix of the probing signals, where convex optimization can be employed. Aiming for generating a beampattern with a desired 3dB main-beam width, an optimization problem was proposed by Stoica et al. [40] as follows

$$\min_{t,\mathbf{R}} \quad -t \tag{8a}$$

$$\text{s.t.} \quad \mathbf{a}^H(\phi_0)\mathbf{R}\mathbf{a}(\phi_0) - \mathbf{a}^H(\phi_m)\mathbf{R}\mathbf{a}(\phi_m) \geq t, \quad \forall \phi_m \in \Omega, \tag{8b}$$

$$\mathbf{a}^H(\phi_1)\mathbf{R}\mathbf{a}(\phi_1) \geq \mathbf{a}^H(\phi_0)\mathbf{R}\mathbf{a}(\phi_0)/2, \tag{8c}$$

$$\mathbf{a}^H(\phi_2)\mathbf{R}\mathbf{a}(\phi_2) \geq \mathbf{a}^H(\phi_0)\mathbf{R}\mathbf{a}(\phi_0)/2, \tag{8d}$$

$$\mathbf{a}^H(\phi_{m+1})\mathbf{R}\mathbf{a}(\phi_{m+1}) \geq \mathbf{a}^H(\phi_m)\mathbf{R}\mathbf{a}(\phi_m),$$
$$\forall \phi_m \in [\phi_1, \phi_0], \tag{8e}$$

$$\mathbf{a}^H(\phi_m)\mathbf{R}\mathbf{a}(\phi_m) \geq \mathbf{a}^H(\phi_{m+1})\mathbf{R}\mathbf{a}(\phi_{m+1}),$$
$$\forall \phi_m \in [\phi_0, \phi_2], \tag{8f}$$

$$\mathbf{R} \succeq 0, \mathbf{R} = \mathbf{R}^H, \tag{8g}$$

$$\text{diag}(\mathbf{R}) = \frac{P_0 \mathbf{1}}{N}, \tag{8h}$$

where $\{\phi_m\}_{m=1}^{M}$ is defined as a fine angular grid that covers the detection angle range of $[-\pi/2, \pi/2]$, $\phi_0$ is the center of the main-beam, $(\phi_2 - \phi_1)$ controls the 3dB width of the main-beam, and $\Omega$ represents the sidelobe region of the beampattern. The vector, $\mathbf{a}(\phi_m) = \left[1, e^{j2\pi\Delta\sin(\phi_m)}, \ldots, e^{j2\pi(N-1)\Delta\sin(\phi_m)}\right] \in \mathbb{C}^{N\times 1}$, is the steering vector of the transmit antenna array with $\Delta$ being the spacing between adjacent antennas normalized by the wavelength, $N$ is the number of antennas of the array, $\mathbf{R}$ is the waveform covariance matrix, $P_0$ is the power budget, and $\mathbf{1}$ is defined as $\mathbf{1} = [1, 1, \ldots, 1]^T \in \mathbb{R}^{N\times 1}$. The constraint (8h) is set to ensure that all antennas transmit the same average power, while (8e) and (8f) are imposed to guarantee that no notches can exist within the main-lobe range. Problem (8a) is a convex semi-definite program, which can be efficiently solved using CVX.

### B. Joint radar-communication problem formulation

Now, we present the IRS assisted ISAC problem formulation in which the BS active beamforming vectors, $\mathbf{w}_k$, are jointly optimized with the IRS passive beamforming vector $\mathbf{v}$. In this ISAC system, the BS antennas are shared among both the radar and the communication services, i.e., the BS transmits a signal that carries information to the communication users, and in the same time, has the characteristics of the desired radar beampattern. Hence, the objective of the ISAC optimization problem is to minimize the Frobenius distance between the covariance matrix of the antenna array of the BS and the desired covariance matrix of the radar beampattern. This objective is optimized given the constraint of meeting the QoS requirements at the communication users as follows,

$$\min_{\mathbf{w}_k, \mathbf{v}} \quad \left\| \sum_{i=1}^{K} \mathbf{w}_i \mathbf{w}_i^H - \mathbf{R} \right\|_F^2 \quad (9a)$$

$$\text{s.t. } \gamma_k \geq \Gamma_k, \quad \forall k, \quad (9b)$$

$$\sum_{i=1}^{K} \|\mathbf{w}_i\|^2 = P_0, \quad (9c)$$

$$|v_l| = 1, \quad l = 1, 2, \cdots, L, \quad (9d)$$

where $\mathbf{R}$ is obtained by solving (8a). Problem (9) is non-convex since the optimization variables $\{\mathbf{w}_k\}_{k=1}^{K}$ and $\mathbf{v}$ are coupled together and multiplied by each others in the SINR constraint in (9b). Therefore, the AO technique is used to tackle this problem by dividing (9) into two sub-problems and solving them iteratively until convergence. In the $n$th iteration, the optimization is done w.r.t. $\mathbf{w}_i$ given $\mathbf{v}^{(n-1)}$, obtained from the previous iteration. Then, the problem is solved w.r.t. $\mathbf{v}$ given the obtained result, $\mathbf{w}_i^{(n)}$, to get $\mathbf{v}^{(n)}$, and so on. In the next section, we illustrate the details of the two sub-problems and how they are solved in each iteration of the AO algorithm.

## IV. PROPOSED SOLUTION FOR PERFECT CSI CASE

In this section, we propose efficient and low complexity solutions to the active and passive beamforming sub-problems of the AO algorithm, when perfect channel state information (CSI) knowledge is available at the BS. Specifically, FP is employed to deal with the SINR constraints in (9b). Then, the active beamforming at BS sub-problem is solved using SDP, while the sub-problem of optimizing the IRS reflection coefficients vector, $\mathbf{v}$, is solved by utilizing SOCP and convex-concave procedure (CCP). We will also show that the proposed SOCP-CCP approach to optimize the IRS vector has much lower complexity than the SDR technique which is widely used in the IRS literature.

### A. Proposed FP-based SDP-SOCP algorithm

Firstly, we need to deal with the fractional SINR constraints in (9b). Therefore, the following rule is used from fractional programming theory [41]

$$\text{Tr}\{\mathbf{A}^H \mathbf{B}^{-1} \mathbf{A}\} = \max_{\mathbf{U}} \text{Tr}\{\mathbf{U}^H \mathbf{A} + \mathbf{A}^H \mathbf{U} - \mathbf{U}^H \mathbf{B} \mathbf{U}\}, \quad (10)$$

Let $f(\mathbf{U}) = \text{tr}\left(\mathbf{U}^H \mathbf{A} + \mathbf{A}^H \mathbf{U} - \mathbf{U}^H \mathbf{B} \mathbf{U}\right)$. Since $f(U)$ is concave in terms of $\mathbf{U}$, the optimal value of $f(\mathbf{U})$ is obtained by letting $\frac{\partial f(\mathbf{U})}{\partial U} = 0$. Therefore, the optimal $\mathbf{U}_{opt}$ is $\mathbf{B}^{-1}\mathbf{A}$. By substituting $\mathbf{U}_{opt}$ into $f(\mathbf{U})$, (10) can be obtained. Using (10) and introducing auxiliary variables $u_k \in \mathbb{C}^{1 \times 1}, \forall k \in \mathcal{K}$, the constraints in (9b) can be transformed into

$$\max_{u_k} \quad 2\mathfrak{Re}\left\{u_k^H \left(\mathbf{h}_k^T + \mathbf{v}^T \text{diag}(\mathbf{r}_k)\mathbf{G}\right) \mathbf{w}_k\right\} - u_k^H \beta_k u_k \geq \Gamma_k,$$
$$\forall \, k \in \mathcal{K}, \quad (11)$$

where $\beta_k = \sum_{\substack{i=1 \\ i \neq k}}^{K} \left|\left(\mathbf{h}_k^T + \mathbf{v}^T \text{diag}(\mathbf{r}_k)\mathbf{G}\right) \mathbf{w}_i\right|^2 + \sigma_n^2$. Therefore, problem (9) can be reformulated as

$$\min_{\mathbf{w}_k, \mathbf{v}, u_k} \quad \left\| \sum_{i=1}^{K} \mathbf{w}_i \mathbf{w}_i^H - \mathbf{R} \right\|_F^2 \quad (12a)$$

$$\text{s.t. } (11), (9c), (9d). \quad (12b)$$

It is difficult to optimize the optimization variables $\mathbf{w}_k$ and $\mathbf{v}$ simultaneously since they are coupled in (11). Hence, the AO technique is adopted to solve the sub-problems corresponding to different sets of variables iteratively until convergence.

*1) Update $\mathbf{w}_k$:* In this sub-problem of the AO algorithm, we solve for $\mathbf{w}_k$, in the $(n+1)$-th iteration, given the IRS reflection coefficients vector, $\mathbf{v}^{(n)}$, and the multiplier, $u_k^{(n)}$, from the previous iteration. Although the constraints (11) and (9c) are convex w.r.t. $\mathbf{w}_k$, problem (12) is not convex because its objective function (12a) is not convex in $\mathbf{w}_k$. This is because the outer product $\mathbf{w}_i \mathbf{w}_i^H$ in (12a) is not convex, so the optimization problem shall be converted to a convex form to be solved using one of the available convex optimization tools. This transformation can be done by converting the active beamforming sub-problem into a SDP problem by letting $\mathbf{W}_k = \mathbf{w}_k \mathbf{w}_k^H$, then performing the optimization over the new variables $\mathbf{W}_k$. Finally, the precoding vectors at BS, $\mathbf{w}_k$, can be extracted from $\mathbf{W}_k$ using Cholesky decomposition after solving the optimization problem.

Now, the constraint (11) shall be converted to be a function of the new optimization variables, $\mathbf{W}_k$. By letting $\mathbf{a}_k^{T,(n)} = \mathbf{h}_k^T + \mathbf{v}^{T,(n)} \text{diag}(\mathbf{r}_k)\mathbf{G}$, constraint (11) can be expressed as

$$\mathfrak{Re}\left\{ 2u_k^{H,(n)} \mathbf{a}_k^{T,(n)} \mathbf{w}_k + |u_k^{(n)}|^2 \left|\mathbf{a}_k^{T,(n)} \mathbf{w}_k\right|^2 \right.$$
$$\left. - |u_k^{(n)}|^2 \left(\sigma^2 + \sum_{i=1}^{K} \left|\mathbf{a}_k^{T,(n)} \mathbf{w}_i\right|^2\right) \right\} \geq \Gamma_k, \, \forall k \in \mathcal{K}. \quad (13)$$

The first two terms of (13) can be expressed as

$$2u_k^{H,(n)}\mathbf{a}_k^{T,(n)}\mathbf{w}_k + |u_k^{(n)}|^2\left|\mathbf{a}_k^{T,(n)}\mathbf{w}_k\right|^2$$
$$= 2u_k^{H,(n)}\mathbf{a}_k^{T,(n)}\mathbf{w}_k + |u_k^{(n)}|^2\left|\mathbf{a}_k^{T,(n)}\mathbf{w}_k\right|^2$$
$$= 2u_k^{H,(n)}\mathbf{a}_k^{T,(n)}\mathbf{w}_k + |u_k^{(n)}|^2\mathbf{w}_k^H\mathbf{a}_k^{*,(n)}\mathbf{a}_k^{T,(n)}\mathbf{w}_k$$
$$= \widetilde{\mathbf{w}}_k^H \mathbf{B}_k^{(n)} \widetilde{\mathbf{w}}_k, \quad (14)$$

where $\widetilde{\mathbf{w}}_k$ and $\mathbf{B}_k^{(n)}$ are given as

$$\widetilde{\mathbf{w}}_k = \begin{bmatrix} \mathbf{w}_k \\ 1 \end{bmatrix}, \text{ and } \mathbf{B}_k^{(n)} = \begin{bmatrix} |u_k^{(n)}|^2 \mathbf{a}_k^{*,(n)} \mathbf{a}_k^{T,(n)} & 0 \\ 2u_k^{H,(n)} \mathbf{a}_k^{T,(n)} & 0 \end{bmatrix}. \quad (15)$$

By letting $\widetilde{\mathbf{a}}_k^{(n)} = [\mathbf{a}_k^{T,(n)}\ 0]^T$, the interference terms under the sum in (13) can be converted to be functions of the new optimization vector, $\widetilde{\mathbf{w}}_k$, as

$$\left|\mathbf{a}_k^{T,(n)}\mathbf{w}_i\right|^2 = \left|\widetilde{\mathbf{a}}_k^{T,(n)}\widetilde{\mathbf{w}}_i\right|^2 = \widetilde{\mathbf{w}}_i^H \mathbf{A}_k^{(n)} \widetilde{\mathbf{w}}_i, \quad (16)$$

where $\mathbf{A}_k^{(n)} = \widetilde{\mathbf{a}}_k^{*,(n)} \widetilde{\mathbf{a}}_k^{T,(n)}$.

Knowing that $\text{Tr}\{\mathbf{XY}\} = \text{Tr}\{\mathbf{YX}\}$, we can derive that

$$\widetilde{\mathbf{w}}_k^H \mathbf{B}_k^{(n)} \widetilde{\mathbf{w}}_k = \text{Tr}\left\{\widetilde{\mathbf{w}}_k^H \mathbf{B}_k^{(n)} \widetilde{\mathbf{w}}_k\right\} = \text{Tr}\left\{\mathbf{B}_k^{(n)} \widetilde{\mathbf{w}}_k \widetilde{\mathbf{w}}_k^H\right\}. \quad (17)$$

By letting $\widehat{\mathbf{W}}_k = \widetilde{\mathbf{w}}_k \widetilde{\mathbf{w}}_k^H \in \mathbb{C}^{L+1 \times L+1}$, the constraints in (13) can be finally transformed to

$$\mathfrak{Re}\left\{\text{Tr}\left\{\mathbf{B}_k^{(n)}\widehat{\mathbf{W}}_k\right\} - |u_k^{(n)}|^2\left(\sigma^2 + \sum_{i=1}^{K} \text{Tr}\left\{\mathbf{A}_k^{(n)}\widehat{\mathbf{W}}_i\right\}\right)\right\}$$
$$\geq \Gamma_k, \quad \forall k \in \mathcal{K}. \quad (18)$$

Finally, the active beamforming SDP sub-problem at BS can be written as

$$\min_{\widehat{\mathbf{W}}_k} \left\|\sum_{i=1}^{K} \widehat{\mathbf{W}}_i(1{:}N, 1{:}N) - \mathbf{R}\right\|_F^2 \quad (19a)$$

$$\text{s.t. } (18) \quad (19b)$$

$$\text{diag}\left(\sum_{i=1}^{K}\widehat{\mathbf{W}}_i(1{:}N, 1{:}N)\right) = \frac{P_0 \mathbf{1}}{N}, \quad (19c)$$

$$\widehat{\mathbf{W}}_k \succeq 0, \widehat{\mathbf{W}}_k = \widehat{\mathbf{W}}_k^H, \text{rank}\left(\widehat{\mathbf{W}}_k\right) = 1, \quad \forall k. \quad (19d)$$

Obviously, all the constraints of (19) are convex except the rank-one constraints in (19d), which is non-convex. By relaxing these rank-one constraints, problem (19) becomes convex and can be solved using standard convex optimization tools such as CVX. Assuming that the optimal solution obtained from solving (19), after dropping the rank-one constraints, is $\widehat{\mathbf{W}}_k^\star$, then we need to extract the precoding vectors at BS, $\mathbf{w}_k^\star$, from $\widehat{\mathbf{W}}_k^\star$. The rank of $\widehat{\mathbf{W}}_k^\star$ is generally greater than 1, thus we will always need to reduce our solution to a rank one matrix. An effective way to find a good rank one approximation is to select the eigenvector of $\widehat{\mathbf{W}}_k^\star$ that corresponds to its maximum eigenvalue [42]. Assuming that the maximum eigenvalue of $\widehat{\mathbf{W}}_k^\star$ is $\lambda_{k,1}$ and the corresponding eigenvector is $\mathbf{q}_{k,1}$, then $\widetilde{\mathbf{w}}_k = \sqrt{\lambda_{k,1}}\mathbf{q}_{k,1}$ can be considered a solution to $\widetilde{\mathbf{w}}_k$. Finally, we can get the precoding vectors at BS as $\mathbf{w}_k^\star = \widetilde{\mathbf{w}}_k^\star(1{:}N)/\widetilde{\mathbf{w}}_k^\star(N+1)$.

*2) Update* $\mathbf{v}$: In this step of the $n$th iteration, we solve for $\mathbf{v}^{(n)}$ given the obtained $\mathbf{w}_k^{(n)}$ from the previous step. The subproblem of problem (12) corresponding to $\mathbf{v}$ is a feasibility-check problem. For improving the convergence of getting a solution for $\mathbf{v}$ for this feasibility-check problem, slack variables $\mathbf{t} = [t_1, \ldots, t_K]^T$ are introduced, which represent the SINR residual of the users. Therefore, the feasibility-check problem of $\mathbf{v}$ can be expressed as follows

$$\max_{\mathbf{v},\mathbf{t}} \|\mathbf{t}\|_1 \quad (20a)$$

$$\text{s.t. } 2\mathfrak{Re}\left\{u_k^{H,(n)}(\mathbf{h}_k^T + \mathbf{v}^T\text{diag}(\mathbf{r}_k)\mathbf{G})\mathbf{w}_k^{(n+1)}\right\} - |u_k^{(n)}|^2$$
$$\left(\sigma_n^2 + \sum_{\substack{i=1 \\ i \neq k}}^{K}\left|(\mathbf{h}_k^T + \mathbf{v}^T\text{diag}(\mathbf{r}_k)\mathbf{G})\mathbf{w}_i^{(n+1)}\right|^2\right) - t_k \geq \Gamma_k,$$
$$\quad (20b)$$

$$t_k \geq 0, \quad \forall k, \quad (20c)$$

$$(9d). \quad (20d)$$

By using $|x|^2 = xx^H$ to expand the interference terms under the summation in (20b), and by applying some mathematical manipulations, the constraints in (20b) can be reformulated as

$$2\mathfrak{Re}\left\{\boldsymbol{\mu}_k^H \mathbf{v}^*\right\} + \mathbf{v}^T \mathbf{M}_k \mathbf{v}^* \leq c_k, \quad k \in \mathcal{K}, \quad (21)$$

where $\boldsymbol{\mu}_k$, $\mathbf{M}_k$, and $c_k$ are given as

$$\boldsymbol{\mu}_k = |u_k^{(n)}|^2 \text{diag}(\mathbf{r}_k)\mathbf{G}\boldsymbol{\Sigma}_k \mathbf{h}_k^* - u_k^{H,(n)}\text{diag}(\mathbf{r}_k)\mathbf{G}\mathbf{w}_k^{(n+1)}, \quad (22)$$

$$\boldsymbol{\Sigma}_k = \sum_{i=1, i\neq k}^{K} \mathbf{w}_i^{(n+1)}\mathbf{w}_i^{H,(n+1)} \quad (23)$$

$$\mathbf{M}_k = |u_k^{(n)}|^2 \text{diag}(\mathbf{r}_k)\mathbf{G}\boldsymbol{\Sigma}_k \mathbf{G}^H \text{diag}(\mathbf{r}_k)^H \quad (24)$$

$$c_k = 2\mathfrak{Re}\left\{u_k^{H,(n)}\mathbf{h}_k^T\mathbf{w}_k^{(n+1)}\right\} - \Gamma_k - t_k - |u_k^{(n)}|^2\left(\sigma_n^2 + \mathbf{h}_k^T\boldsymbol{\Sigma}_k \mathbf{h}_k^*\right). \quad (25)$$

Thus, the constraints in (21) are quadratic constraints, and they can be equivalently transformed into the standard convex second order cone (SOC) constraints. Thus, the sub-problem for updating the IRS coefficients, $\mathbf{v}$, can be written as

$$\max_{\mathbf{v},\mathbf{t}} \|\mathbf{t}\|_1 \quad (26a)$$

$$\text{s.t. } \left\|\begin{array}{c}\mathbf{M}_k^{\frac{1}{2}}\mathbf{v}^* \\ \frac{1-c_k}{2} + \mathfrak{Re}\left\{\boldsymbol{\mu}_k^H \mathbf{v}^*\right\}\end{array}\right\|_2 \leq \frac{1+c_k}{2} - \mathfrak{Re}\left\{\boldsymbol{\mu}_k^H \mathbf{v}^*\right\},$$
$$\quad (26b)$$

$$k \in \mathcal{K},$$
$$(20c), (9d). \quad (26c)$$

Although the cost function of (26) is linear, and the constraints in (26b) are SOC constraints, problem (26) cannot be considered a convex SOCP due to the non-convex unit-modulus constraint (9d). To handle this, the penalty CCP method [43] is used to find a feasible solution that satisfies both SINR constraints and the unit-modulus constraints. Specifically, the unit-modulus constraints $|v_l|^2 = 1$ can be equivalently expressed as $1 \leq |v_l|^2 \leq 1$, where the non-convex parts of these inequalities are linearized as $|v_l^{[j]}|^2 - 2\mathfrak{Re}(v_l^H v_l^{[j]}) \leq -1$. Then, the slack variables $\boldsymbol{\xi} = [\xi_1, \ldots, \xi_{2L}]^T$ are imposed following the CCP procedure over the equivalent constraints, which allow us to write the equivalent problem as

$$\max_{\mathbf{v},\mathbf{t},\boldsymbol{\xi}} \quad \|\mathbf{t}\|_1 - \rho^{[j]} \|\boldsymbol{\xi}\|_1 \tag{27a}$$

$$\text{s.t.} \quad (26b), (20c), \tag{27b}$$

$$|v_l^{[j]}|^2 - 2\Re(v_l^H v_l^{[j]}) \le \xi_l - 1, \quad 1 \le l \le L \tag{27c}$$

$$|v_l|^2 \le 1 + \xi_{L+l}, \quad 1 \le l \le L \tag{27d}$$

$$\boldsymbol{\xi} \ge 0, \tag{27e}$$

where $\rho^{[j]}$ is a regularization factor adjusting the influence of the penalty term $\|\boldsymbol{\xi}\|_1$, that controls the feasibility of the constraints. Problem (27) is a SOCP which can be efficiently solved using the CVX tool. **Algorithm 1** summarizes the CCP procedure for finding a feasible solution for $\mathbf{v}$. To prevent numerical issues, we enforce an upper limit $\rho_{max}$ to avoid convergence problems, as an excessively large $\rho^{[j]}$ might hinder finding a feasible solution during iterations. The convergence of **Algorithm 1** is monitored by the stopping criterion $\|\mathbf{v}^{[j]} - \mathbf{v}^{[j-1]}\|_1 \le \nu_1$, where $\nu_1$ is a small positive value. Additionally, the stopping criterion $\|\boldsymbol{\xi}\|_1 \le \nu_2$ ensures that the unit-modulus constraints of the original problem (26) are satisfied with a sufficiently low $\nu_2$.

---

**Algorithm 1:** CCP optimization algorithm for solving IRS beamforming problem (26)

1 **Initialize:** Starting feasible point $\mathbf{v}^{[0]}$, $\rho^{[0]}$, $\tau \ge 1$, and set $j = 0$.
2 **while** $j \le J_{max}$ **do**
3  | Update $\mathbf{v}^{[j+1]}$ by solving problem (27);
4  | $\rho^{[j+1]} = \min\{\tau \rho^{[j]}, \rho_{max}\}$;
5  | $j = j + 1$;
6  | **if** $(\|\mathbf{v}^{[j]} - \mathbf{v}^{[j-1]}\|_1 \le \nu_1$ and $\|\boldsymbol{\xi}\|_1 \le \nu_2)$ **then**
7  |  | End while loop;
8  |  | Return $\mathbf{v}^{(n+1)} = \mathbf{v}^{[j]}$;
9  | **end**
10 **end**

---

*3) Update $u_k$:* In this step during the $n$th iteration, the elements $u_k$ are updated based on (10) as

$$u_k^{(n+1)} = \frac{\left(\mathbf{h}_k^T + \mathbf{v}^{T,(n+1)} \text{diag}(\mathbf{r}_k)\mathbf{G}\right) \mathbf{w}_k^{(n+1)}}{\sum\limits_{\substack{i=1 \\ i \ne k}}^{K} \left|\left(\mathbf{h}_k^T + \mathbf{v}^{T,(n+1)} \text{diag}(\mathbf{r}_k)\mathbf{G}\right) \mathbf{w}_i^{(n+1)}\right|^2 + \sigma_n^2}. \tag{28}$$

Finally, the steps of the proposed FP-SDP-SOCP based AO beamforming algorithm are summarized in **Algorithm 2**.

### B. Optimizing active and passive beamforming using SDR

In this subsection, we briefly present a benchmark solution to problem (9) to compare with the proposed FP-SDP-SOCP based algorithm in the previous subsection. Both the active and passive beamforming sub-problems of the original problem (9) can be converted to a SDP, then solved iteratively using SDR. However, the complexity of this benchmark is much higher than the proposed FP-SDP-SOCP based solution when

---

**Algorithm 2:** Proposed FP-SDP-SOCP based AO algorithm for solving (9)

1 **Initialize:** Initialize a starting feasible point $\mathbf{v}^{(0)}$, $\mathbf{w}_k^{(0)}$ and $u_k^{(0)}$ and set $n = 0$.
2 **repeat**
3  | Update $\mathbf{w}_k^{(n+1)}$ by solving problem (19), given $\mathbf{v}^{(n)}$ and $u_k^{(n)}$;
4  | Update $\mathbf{v}^{(n+1)}$ by solving problem (27) using Algorithm 1, given $\mathbf{w}_k^{(n+1)}$ and $u_k^{(n)}$;
5  | Update $u_k^{(n+1)}$ using (28) given $\mathbf{v}^{(n+1)}$ and $\mathbf{w}_k^{(n+1)}$;
6  | Increment the iteration number, $n = n + 1$;
7 **until** $\Xi^{(n+1)} - \Xi^{(n)} \le \nu_3$, where $\Xi^{(n)} = \left\| \sum_{i=1}^{K} \mathbf{w}_i^{(n)} \mathbf{w}_i^{H,(n)} - \mathbf{R} \right\|_F^2$;

---

optimizing the IRS reflection coefficients. The reason for this is that the bi-SDP benchmark uses the SDP technique to optimize the IRS vector, whose complexity explodes specially when the number of IRS reflectors, $L$, becomes large.

By converting (9) into a SDP, both the active and reflective beamforming sub-problems are transformed to convex problems, which can be solved using CVX. The two convex SDP sub-problems are then solved iteratively using AO until convergence. This transformation is done as follows. The intended signal power of user $k$ can be written as

$$\left|\left(\mathbf{h}_k^T + \mathbf{v}^T \text{diag}(\mathbf{r}_k)\mathbf{G}\right) \mathbf{w}_k\right|^2 = \left|\widetilde{\mathbf{v}}^T \mathbf{H}_k \mathbf{w}_k\right|^2, \tag{29}$$

where $\mathbf{H}_k = \begin{bmatrix} \text{diag}(\mathbf{r}_k)\mathbf{G} \\ \mathbf{h}_k^T \end{bmatrix}$ and $\widetilde{\mathbf{v}} = \begin{bmatrix} \mathbf{v} \\ 1 \end{bmatrix}$. Hence, (29) can be expanded as

$$\left|\widetilde{\mathbf{v}}^T \mathbf{H}_k \mathbf{w}_k\right|^2 = \mathbf{w}_k^H \mathbf{H}_k^H \widetilde{\mathbf{v}}^* \widetilde{\mathbf{v}}^T \mathbf{H}_k \mathbf{w}_k. \tag{30}$$

Given $\text{Tr}\{\mathbf{AB}\} = \text{Tr}\{\mathbf{BA}\}$, and $\mathbf{W}_k = \mathbf{w}_k \mathbf{w}_k^H \in \mathbb{C}^{N \times N}$ and $\widetilde{\mathbf{V}} = \widetilde{\mathbf{v}}^* \widetilde{\mathbf{v}}^T \in \mathbb{C}^{L+1 \times L+1}$, (30) can be expressed as

$$\left|\widetilde{\mathbf{v}}^T \mathbf{H}_k \mathbf{w}_k\right|^2 = \text{Tr}\{\mathbf{H}_k^H \widetilde{\mathbf{v}}^* \widetilde{\mathbf{v}}^T \mathbf{H}_k \mathbf{W}_k\} = \text{Tr}\{\mathbf{H}_k \mathbf{w}_k \mathbf{w}_k^H \mathbf{H}_k^H \widetilde{\mathbf{V}}\}. \tag{31}$$

Both $\mathbf{W}_k$ and $\widetilde{\mathbf{V}}$ are symmetric positive semi-definite rank-one matrices, since they are the outer products of $\mathbf{w}_k$ and $\widetilde{\mathbf{v}}$, respectively. Therefore, the SDP of the active beamforming at BS sub-problem, given the reflective beamforming vector,

$\widetilde{\mathbf{v}}^{(n)}$, can be given as

$$\max_{\mathbf{W}_k} \left\| \sum_{i=1}^{K} \mathbf{W}_i - \mathbf{R} \right\|_F^2 \tag{32a}$$

$$\text{s.t.} \quad \text{Tr}\left\{\mathbf{H}_k^H \widetilde{\mathbf{v}}^{*,(n)} \widetilde{\mathbf{v}}^{T,(n)} \mathbf{H}_k \mathbf{W}_k\right\} - \Gamma_k \left(\sigma_n^2\right.$$
$$\left. + \sum_{\substack{i=1\\i\neq k}}^{K} \text{Tr}\left\{\mathbf{H}_k^H \widetilde{\mathbf{v}}^{*,(n)} \widetilde{\mathbf{v}}^{T,(n)} \mathbf{H}_k \mathbf{W}_i\right\}\right) \geq 0, \quad \forall k, \tag{32b}$$

$$\text{diag}\left(\sum_{i=1}^{K} \mathbf{W}_i\right) = \frac{P_0 \mathbf{1}}{N}, \tag{32c}$$

$$\mathbf{W}_k \succeq \mathbf{0}, \quad \text{rank}(\mathbf{W}_k) = 1, \quad \forall k. \tag{32d}$$

Also, the reflective beamforming sub-problem at IRS, given the active beamforming vectors, $\mathbf{w}_k^{(n+1)}$, can be given as

$$\max_{\widetilde{\mathbf{V}}, t_k} \sum_{i=1}^{K} t_i \tag{33a}$$

$$\text{s.t.} \quad \text{Tr}\left\{\mathbf{H}_k \mathbf{w}_k^{(n+1)} \mathbf{w}_k^{H,(n+1)} \mathbf{H}_k^H \widetilde{\mathbf{V}}\right\} - \Gamma_k \left(\sigma_n^2 + \right.$$
$$\left. \sum_{\substack{i=1\\i\neq k}}^{K} \text{Tr}\left\{\mathbf{H}_k \mathbf{w}_i^{(n+1)} \mathbf{w}_i^{H,(n+1)} \mathbf{H}_k^H \widetilde{\mathbf{V}}\right\}\right) - t_k \geq 0, \quad \forall k, \tag{33b}$$

$$t_k \geq 0, \tag{33c}$$

$$\widetilde{\mathbf{V}}(i,i) = 1, \quad i = 1, 2, \cdots, L+1, \tag{33d}$$

$$\widetilde{\mathbf{V}} \succeq \mathbf{0}, \quad \text{rank}(\widetilde{\mathbf{V}}) = 1. \tag{33e}$$

Finally, assuming that the optimal converged solution obtained by iteratively solving (32) and (33), after dropping the rank-one constraints, is $\mathbf{W}_k^\star$ and $\widetilde{\mathbf{V}}^\star$, then the precoding vectors $\mathbf{w}_k^\star$ and $\widetilde{\mathbf{v}}^\star$ can be extracted using the eigenvalue decomposition discussed earlier. Nonetheless, the reflective beamforming vector at IRS is computed as $\mathbf{v}^\star = \widetilde{\mathbf{v}}^\star(1{:}L)/\widetilde{\mathbf{v}}^\star(L+1)$.

*C. Complexity analysis*

Each iteration of the alternation optimization algorithm consists of two sub-problems; the active beamforming sub-problem in (19) and the passive beamforming one in (27). Now, we compare the complexity of the proposed FP-SDP-SOCP algorithm against the complexity of the SDR benchmark to show the significance of the proposed algorithm.

Starting with the active beamforming sub-problem, the complexity of either (19) in the proposed algorithm or (32) in the benchmark is the same since the two problems are based on SDP. Then, the complexity of both problems is determined by the complexity of the interior point method used for solving the SDP, which is given as $\mathcal{O}(\max(N,K)^4 N^{1/2} \log(1/\zeta))$ [44], where $\zeta$ is the accuracy of the solution. It should be noted that although there is no complexity advantage in the active beamforming sub-problem for the proposed algorithm over the benchmark, this is of less significance since the number of BS antennas $N$ is typically much lower than the number of IRS reflection elements. Consequently, the passive beamforming sub-problem is of more importance to reduce its complexity since it is the higher dimensional problem, in terms of the number of optimization variables.

Now, we analyze the complexity of the passive beamforming sub-problem at IRS which becomes more critical as the number of IRS elements goes large. We compare the complexity of both the benchmark sub-problem in (33) against the proposed algorithm sub-problem in (27) as follows.

1) The complexity of the benchmark sub-problem in (33) depends on the complexity of the interior point method since it is a SDP. Therefore, the complexity of (33) is given as $\mathcal{O}\left(\max(L+1,K)^4 (L+1)^{1/2} \log(1/\zeta)\right)$ [44], where $\zeta$ is the accuracy of the solution.

2) However, the passive beamforming sub-problem of the proposed algorithm in (27) is formulated into a SOCP not SDP, which makes its complexity analysis different. Therefore, the worst case complexity is determined by the SOCP in each step. The SOCP is solved using interior-point methods and its complexity depends on the number of non-linear constraints, variables and the dimension of each second order cone (SOC) constraint [45]. By ignoring the linear constraints, the number of non-linear constraints in (27) is $K$. Hence, it requires $\mathcal{O}\left(\sqrt{K} \log \frac{1}{\zeta}\right)$ iterations to converge with $\zeta$ accuracy when the algorithm terminates. In each iteration, the number of arithmetic operation is at most $\mathcal{O}\left((L+K)^2(LK+k^2)\right)$ [45].

In summary, the SOCP approach exhibits significantly improved worst-case complexity compared to the SDR benchmark. Unlike the semi-definite formulation, there is no requirement to introduce the additional matrix $\widetilde{\mathbf{V}}$ in the benchmark, leading to a substantial reduction in the number of variables involved in the resulting optimization process.

## V. ROBUST BEAMFORMING FOR CSI WITH UNCERTAINTY

In this section, a robust solution is proposed for the joint beamforming problem when the channel has uncertainty. Since the IRS-users channel vectors, $\mathbf{r}_k$, are more challenging to estimate than the BS-IRS and the BS-users channels, we assume that the estimates of $\mathbf{r}_k$ are imperfect at BS. The IRS-users channel vectors, $\mathbf{r}_k$, can be modeled as $\mathbf{r}_k = \widehat{\mathbf{r}}_k + \boldsymbol{\Delta}_k, \forall k$, where $\widehat{\mathbf{r}}_k$ represent the erroneously estimated channel vectors at BS, and $\boldsymbol{\Delta}_k, \forall k$ represent the complementary error vectors. In this section, a bounded channel error model [46] is utilized, which imposes a limit on the magnitude of the error vectors, i.e., $\|\boldsymbol{\Delta}_k\|_2 \leq \varepsilon_k, \forall k$, where $\varepsilon_k$ is the radius of the uncertainty ball known by the BS. Therefore, in the presence of imperfect CSI at BS, the worst-case robust ISAC design problem is

formulated as

$$\min_{\mathbf{w}_k,\mathbf{v},u_k} \left\| \sum_{i=1}^{K} \mathbf{w}_i \mathbf{w}_i^H - \mathbf{R} \right\|_F^2 \quad (34a)$$

s.t. $\max_{u_k} 2\Re\{u_k^H (\mathbf{h}_k^T + (\widehat{\mathbf{r}}_k + \boldsymbol{\Delta}_k)^T \boldsymbol{\Theta} \mathbf{G}) \mathbf{w}_k\} - |u_k|^2 \Big( \sigma_n^2$

$$+ \sum_{\substack{i=1 \\ i \neq k}}^{K} \left| (\mathbf{h}_k^T + (\widehat{\mathbf{r}}_k + \boldsymbol{\Delta}_k)^T \boldsymbol{\Theta} \mathbf{G}) \mathbf{w}_i \right|^2 \Big) \geq \Gamma_k, \quad (34b)$$

$$\forall \|\boldsymbol{\Delta}_k\|_2 \leq \varepsilon_k, \quad \forall k, \quad (34c)$$

$$(9c), (9d), \quad (34d)$$

where $\boldsymbol{\Theta} = \mathrm{diag}(\mathbf{v})$ is the IRS coefficients diagonal matrix. Again, the AO is used to solve (34) by iteratively updating $\mathbf{w}_k, \mathbf{v}$, and $u_k$ as in the perfect CSI case discussed in Sec. IV. However, we have to deal with the uncertainty vectors, $\boldsymbol{\Delta}_k$, in each sub-problem of AO, which is illustrated in the following.

*A. Robust joint active and passive beamforming algorithm*

In this subsection, we illustrate how $\mathbf{w}_k, \mathbf{v}$, and $u_k$ are updated during each iteration of the AO algorithm in the presence of channel uncertainty, $\boldsymbol{\Delta}_k$. These uncertainty vectors affect the feasibility of the constraints of each sub-problem, hence, the constraints must be satisfied for all possible values of uncertainty vectors within the uncertainty sphere. This can be done by transforming each constraint with uncertainty into a LMI as we explain in the following steps.

*1) Update $\mathbf{w}_k$:* In the $(n+1)$th iteration of the AO algorithm, (34) is solved for $\mathbf{w}_k$ given the IRS coefficients matrix from the previous iteration, $\boldsymbol{\Theta}^{(n)}$. However, and similar to the perfect CSI case, the active beamforming sub-problem at BS must be converted to SDP first, then the new constraints shall be converted to LMIs to consider the uncertainty. Therefore, we consider the QoS constraints in (18), substituting $\mathbf{r}_k$ with $\widehat{\mathbf{r}}_k + \boldsymbol{\Delta}_k$ as follows.

Assume that optimization variables $\widehat{\mathbf{W}}_k$ are represented as

$$\widehat{\mathbf{W}}_k = \begin{bmatrix} \overline{\mathbf{W}}_k \in \mathbb{C}^{N \times N} & \overline{\mathbf{w}}_k \in \mathbb{C}^{N \times 1} \\ \overline{\mathbf{w}}_k^H \in \mathbb{C}^{1 \times N} & 1 \end{bmatrix}, \quad (35)$$

where $\overline{\mathbf{W}}_k = \widehat{\mathbf{W}}_k(1{:}N, 1{:}N)$ and $\overline{\mathbf{w}}_k = \widehat{\mathbf{W}}_k(1{:}N, N{+}1)$. By using the definitions of $\widehat{\mathbf{B}}_k^{(n)}$ and $\mathbf{A}_k^{(n)}$ in (16) and (15), respectively, the trace terms in (18) can be written as

$$\mathrm{Tr}\{\mathbf{B}_k \widehat{\mathbf{W}}_k\} = \mathrm{Tr}\{|u_k|^2 \mathbf{a}_k^* \mathbf{a}_k^T \overline{\mathbf{W}}_k\} + 2u_k^H \mathbf{a}_k^T \overline{\mathbf{w}}_k$$
$$= |u_k|^2 \mathbf{a}_k^T \overline{\mathbf{W}}_k \mathbf{a}_k^* + 2u_k^H \mathbf{a}_k^T \overline{\mathbf{w}}_k, \quad (36)$$

$$\sum_{i=1}^{K} \mathrm{Tr}\{\mathbf{A}_k \widehat{\mathbf{W}}_i\} = \mathrm{Tr}\Big\{\mathbf{A}_k \sum_{i=1}^{K} \widehat{\mathbf{W}}_i\Big\} = \mathrm{Tr}\Big\{\mathbf{a}_k^* \mathbf{a}_k^T \sum_{i=1}^{K} \overline{\mathbf{W}}_i\Big\}$$
$$= \mathbf{a}_k^T \Big(\sum_{i=1}^{K} \overline{\mathbf{W}}_i\Big) \mathbf{a}_k^*. \quad (37)$$

Therefore, by substituting (36) and (37) in (18), the new constraints can be given as

$$u_k^{H,(n)} \mathbf{a}_k^{T,(n)} \overline{\mathbf{w}}_k + u_k^{(n)} \overline{\mathbf{w}}_k^H \mathbf{a}_k^{*,(n)} - |u_k^{(n)}|^2 \Big( \sigma_n^2$$
$$+ \mathbf{a}_k^{T,(n)} \overline{\boldsymbol{\Sigma}}_{-k} \mathbf{a}_k^{*,(n)} \Big) \geq \Gamma_k, \quad \forall k, \quad (38)$$

where $\overline{\boldsymbol{\Sigma}}_{-k} = \sum_{i=1, i \neq k}^{K} \overline{\mathbf{W}}_k$. With the aid of $\mathbf{a}^{T,(n)} = \mathbf{h}_k^T + (\widehat{\mathbf{r}}_k + \boldsymbol{\Delta}_k)^T \boldsymbol{\Theta}^{(n)} \mathbf{G}$, and after performing some mathematical manipulations, (38) can be rearranged in the form

$$\boldsymbol{\Delta}_k^T \mathbf{X}_k \boldsymbol{\Delta}_k^* + \boldsymbol{\Delta}_k^T \mathbf{x}_k + \mathbf{x}_k^H \boldsymbol{\Delta}_k^* + d_k \geq \Gamma_k, \quad \forall k, \quad (39)$$

where,

$$\mathbf{X}_k = -|u_k^{(n)}|^2 \boldsymbol{\Theta}^{(n)} \mathbf{G} \overline{\boldsymbol{\Sigma}}_{-k} \mathbf{G}^H \boldsymbol{\Theta}^{H,(n)} \quad (40)$$

$$\mathbf{x}_k = u_k^{H,(n)} \boldsymbol{\Theta}^{(n)} \mathbf{G} \overline{\mathbf{w}}_k - |u_k^{(n)}|^2 \boldsymbol{\Theta}^{(n)} \mathbf{G} \overline{\boldsymbol{\Sigma}}_{-k} (\mathbf{h}_k^* + \mathbf{G}^H \boldsymbol{\Theta}^{H,(n)} \widehat{\mathbf{r}}_k^*) \quad (41)$$

$$d_k = u_k^{H,(n)} (\mathbf{h}_k^T + \widehat{\mathbf{r}}_k^T \boldsymbol{\Theta}^{(n)} \mathbf{G}) \overline{\mathbf{w}}_k + u_k^{(n)} \overline{\mathbf{w}}_k^H (\mathbf{h}_k^* + \mathbf{G}^H \boldsymbol{\Theta}^{H,(n)} \widehat{\mathbf{r}}_k^*)$$
$$- |u_k^{(n)}|^2 (\mathbf{h}_k^T + \widehat{\mathbf{r}}_k^T \boldsymbol{\Theta}^{(n)} \mathbf{G}) \overline{\boldsymbol{\Sigma}}_{-k} (\mathbf{h}_k^* + \mathbf{G}^H \boldsymbol{\Theta}^{H,(n)} \widehat{\mathbf{r}}_k^*) - |u_k^{(n)}|^2 \sigma_n^2. \quad (42)$$

Using the S-procedure in [47] that is used for robust optimization problems, the constraints in (39) can be transformed into equivalent LMIs as

$$\begin{bmatrix} \kappa_k \mathbf{I}_L + \mathbf{X}_k & \mathbf{x}_k \\ \mathbf{x}_k^H & d_k - \Gamma_k - \kappa_k \varepsilon_k^2 \end{bmatrix} \succeq \mathbf{0}, \quad \forall k, \quad (43)$$

where $\kappa_k$ are slack variables. Finally, the robust active beamforming sub-problem at BS can be given as

$$\min_{\widehat{\mathbf{W}}_k} \left\| \sum_{i=1}^{K} \overline{\mathbf{W}}_i - \mathbf{R} \right\|_F^2 \quad (44a)$$

s.t. $(43), (19c), (19d). \quad (44b)$

*2) Update $\mathbf{v}$:* In this step of the $(n+1)$th iteration, the IRS coefficients matrix, $\boldsymbol{\Theta}$, is optimized given the obtained active beamforming vectors at BS, $\mathbf{w}_k^{(n+1)}$. Thus, problem (20) is solved considering the uncertainty vectors, $\boldsymbol{\Delta}_k$. We shall deal with the constraints in (20b), since they have the uncertainty vectors, and they can be recast as

$$2\Re\left\{ u_k^{H,(n)} \mathbf{w}_k^{T,(n+1)} (\mathbf{h}_k + \mathbf{G}^T \boldsymbol{\Theta} \mathbf{r}_k) \right\} - |u_k^{(n)}|^2$$
$$\left( \sigma_n^2 + \left\| \mathbf{T}_{-k}^{T,(n+1)} (\mathbf{h}_k + \mathbf{G}^T \boldsymbol{\Theta} \mathbf{r}_k) \right\|_2^2 \right) - t_k \geq \Gamma_k, \quad \forall k, \quad (45)$$

where $\mathbf{T}_{-k}^{(n+1)} = [\mathbf{w}_1 \ldots \mathbf{w}_{k-1} \mathbf{w}_{k+1} \ldots \mathbf{w}_K]^{T,(n+1)}$. To express the above inequalities as LMIs, the following lemma [46] is used.

**Lemma 1.** *(Schur's Complement): Let*

$$\mathbf{Z} = \begin{bmatrix} \mathbf{A} & \mathbf{B}^H \\ \mathbf{B} & \mathbf{C} \end{bmatrix}, \quad (46)$$

*be a Hermitian matrix with $\mathbf{C} \succ \mathbf{0}$. Then, $\mathbf{Z} \succeq \mathbf{0}$ if and only if $\boldsymbol{\Delta}_\mathbf{C} \succeq \mathbf{0}$, where $\boldsymbol{\Delta}_\mathbf{C}$ is the Schur complement of $\mathbf{C}$ in $\mathbf{Z}$ and is given by $\boldsymbol{\Delta}_\mathbf{C} = \mathbf{A} - \mathbf{B}^H \mathbf{C}^{-1} \mathbf{B}$.*

By letting $q = u_k^{H,(n)} \mathbf{w}_k^{T,(n+1)} (\mathbf{h}_k + \mathbf{G}^T \boldsymbol{\Theta} \mathbf{r}_k)$ and $\mathbf{b} = [(\mathbf{h}_k^T + \mathbf{r}_k^T \boldsymbol{\Theta} \mathbf{G}) \mathbf{T}_{-k}^{(n+1)} \ \sigma_n]^T$, (45) can be expressed as

$$q + q^H - (u_k^{(n)} \mathbf{b})^H (u_k^{(n)} \mathbf{b}) - a_k - \Gamma_k \geq 0, \quad \forall k. \quad (47)$$

Then, by using Lemma 1, the constraints in (47) can be expressed using LMIs as

$$\begin{bmatrix} q + q^H - t_k - \Gamma_k & u_k^{H,(n)} \mathbf{b}^H \\ u_k^{(n)} \mathbf{b} & \mathbf{I}_K \end{bmatrix} \succeq \mathbf{0}, \quad \forall k. \quad (48)$$

By substituting $(\mathbf{a}_k^{(n)})^T =$ in (48), and after performing some mathematical manipulations, (48) can be written as

$$\begin{bmatrix} z_1+z_1^H-t_k-\Gamma_k+\mathbf{g}^H\boldsymbol{\Delta}_k+\boldsymbol{\Delta}_k^H\mathbf{g} & \mathbf{z}_2^H+\boldsymbol{\Delta}_k^H\mathbf{D}^H & u_k^{H,(n)}\sigma_n \\ \mathbf{z}_2 + \mathbf{D}\boldsymbol{\Delta}_k & \mathbf{I}_{(K-1)} & \mathbf{0}^{K-1\times 1} \\ u_k^{(n)}\sigma_n & \mathbf{0}^{1\times K-1} & 1 \end{bmatrix}$$
$$\succeq \mathbf{0}, \quad \forall k \in \mathcal{K}, \quad \forall \boldsymbol{\Delta}_k : \|\boldsymbol{\Delta}_k\| \leq \varepsilon_k, \quad (49)$$

where,

$$z_1 = u_k^{H,(n)}\mathbf{w}_k^{T,(n+1)}\left(\mathbf{h}_k+\mathbf{G}^T\boldsymbol{\Theta}\widehat{\mathbf{r}}_k\right),$$
$$\mathbf{z}_2 = u_k^{(n)}\mathbf{T}_{-k}^{T,(n+1)}\left(\mathbf{h}_k+\mathbf{G}^T\boldsymbol{\Theta}\widehat{\mathbf{r}}_k\right),$$
$$\mathbf{g} = u_k^{(n)}\boldsymbol{\Theta}^*\mathbf{G}^*\mathbf{w}_k^{*,(n+1)}, \quad \mathbf{D} = u_k^{(n)}\mathbf{T}_{-k}^{T,(n+1)}\mathbf{G}^T\boldsymbol{\Theta}. \quad (50)$$

The constraints in (49) can be written in an equivalent form as

$$\mathbf{Y}(\boldsymbol{\Theta}) \succeq \mathbf{P}^H(\boldsymbol{\Theta})\boldsymbol{\Delta}_k\mathbf{Q}+\mathbf{Q}^H\boldsymbol{\Delta}_k^H\mathbf{P}(\boldsymbol{\Theta}), \quad \forall \boldsymbol{\Delta}_k : \|\boldsymbol{\Delta}_k\| \leq \varepsilon_k, \quad (51)$$

where,

$$\mathbf{Y}(\boldsymbol{\Theta}) = \begin{bmatrix} z_1+z_1^H-t_k-\Gamma_k & \mathbf{z}_2^H & u_k^{H,(n)}\sigma_n \\ \mathbf{z}_2 & \mathbf{I}_{(K-1)} & \mathbf{0}^{K-1\times 1} \\ u_k^{(n)}\sigma_n & \mathbf{0}^{1\times K-1} & 1 \end{bmatrix},$$
$$\mathbf{P}(\boldsymbol{\Theta}) = \begin{bmatrix} \mathbf{g} & \mathbf{D} & \mathbf{0}^{L\times 1} \end{bmatrix},$$
$$\mathbf{Q} = \begin{bmatrix} -1 & \mathbf{0}^{1\times K-1} & 0 \end{bmatrix}. \quad (52)$$

To get LMIs which perfectly satisfy the constraints in (51), the following lemma [46] is used.

**Lemma 2.** *For any matrices* $\mathbf{B}$, $\mathbf{C}$, *and a hermitian matrix,* $\mathbf{A} = \mathbf{A}^H$,

$$\mathbf{A} \succeq \mathbf{B}^H\mathbf{U}\mathbf{C}+\mathbf{C}^H\mathbf{U}^H\mathbf{B}, \quad \forall \mathbf{U} : \|\mathbf{U}\|_2 \leq \varepsilon, \quad (53)$$

*if and only if there exists a* $\delta \geq 0$ *such that*

$$\mathbf{Z} = \begin{bmatrix} \mathbf{A} - \delta\mathbf{C}^H\mathbf{C} & -\varepsilon\mathbf{B}^H \\ -\varepsilon\mathbf{B} & \delta\mathbf{I} \end{bmatrix} \succeq 0. \quad (54)$$

Thus, by applying Lemma 2 to (51), the constraints in (49) can be transformed to their equivalent LMIs using the slack variables, $\delta_k$, as

$$\begin{bmatrix} z_1+z_1^H-t_k-\Gamma_k-\delta_k & \mathbf{z}_2^H & u_k^{H,(n)}\sigma_n & -\varepsilon_k\mathbf{g}^H \\ \mathbf{z}_2 & \mathbf{I}_{(K-1)} & \mathbf{0}^{K-1\times 1} & -\varepsilon_k\mathbf{D}^H \\ u_k^{(n)}\sigma_n & \mathbf{0}^{1\times K-1} & 1 & \mathbf{0}^{1\times L} \\ -\varepsilon_k\mathbf{g} & -\varepsilon_k\mathbf{D} & \mathbf{0}^{L\times 1} & \delta_k\mathbf{I}_L \end{bmatrix}$$
$$\succeq 0, \quad \forall k \in \mathcal{K}. \quad (55)$$

Lastly, we apply the penalty CCP framework in Sec. IV, by imposing the slack variables $\boldsymbol{\xi} = [\xi_1, \ldots, \xi_{2L}]^T$, to convert the non-convex unit-modulus constraints in (9d) to their equivalent convex form. Therefore, the final form of the robust passive beamforming sub-problem at IRS is written as

$$\max_{\substack{\mathbf{v},\mathbf{t},\boldsymbol{\xi} \\ \boldsymbol{\delta}}} \|\mathbf{t}\|_1 - \rho^{[j]}\|\boldsymbol{\xi}\|_1 \quad (56a)$$

$$\text{s.t.} \quad (55), (20c), (27c), (27d), (27e) \quad (56b)$$

$$\delta_k \geq 0. \quad (56c)$$

*3) Update* $u_k$: Now, given $\mathbf{w}_k^{(n+1)}$ and $\boldsymbol{\Theta}^{(n+1)}$, the multiplier $u_k$ is optimized such that it maximizes the parameter, $\delta_k$, which controls the feasibility of the constraints in (55). Hence, $u_k$ can be updated as

$$\max_{\delta_k, u_k} \quad \delta_k \quad (57a)$$

$$\text{s.t.} \quad (55), \quad (57b)$$

$$\delta_k \geq 0. \quad (57c)$$

In conclusion, the proposed robust beamforming algorithm is performed by updating $\mathbf{w}_k$, $\mathbf{v}$ and $u_k$ alternatively using the AO technique, which was illustrated in the presented previous three steps.

## VI. SIMULATION RESULTS

In this section, we present numerical results using Monte Carlo simulations to validate the effectiveness of the proposed beamforming optimization techniques for IRS assisted ISAC systems. The performance of the proposed low complexity FP-SDP-SOCP based algorithm is compared against the benchmark technique presented in Sec. IV-B. Moreover, the performance of the proposed robust beamforming approach is also validated by changing the uncertainty radius of the channels and observing its effect on the performance of the obtained radar-communication beampatterns. We plot the PSLR and MSE as performance measures to assess the quality of the obtained beampatterns. Throughout these simulations, specific parameters are fixed, with total transmit power at BS set to $P_0 = 20$dBm, the number of BS antennas $N = 20$, the number of communication users $K = 4$, and a uniform linear array (ULA) is employed with a half-wavelength spacing, i.e., $d = \lambda/2$, between adjacent antennas.

The simulation parameters used follow [17], [48]. The distance between the BS and the IRS is assumed $d_{IB} = 25\ m$, the IRS-user distance is $20\ m$, and the BS-user distance is $50\ m$. The path loss at the reference distance in (3) is $\eta_0 = 10^{-3}$, the path loss exponents for the IRS-user links and the BS-IRS link are set as $\alpha_{IU} = \alpha_{BI} = 2.2$, while the path loss exponent for the BS-user links is assumed as $\alpha_{BU} = 3.5$ in this simulation setup. The noise power $\sigma_n^2$ is set to be $-114$ dBm, and the Rician factors in (1) and (2) are set to be $K_{IB} = K_{UI} = 2.2$.

To assess the effectiveness of the proposed algorithm, we compare the obtained beampatterns using it against the radar-only beampattern and the beampatterns of the no-IRS scenario. Moreover, we spot the effect of robust beamforming on the obtained beampatterns under different uncertainty radii. We analyze the performance of the different scenarios using their 3dB beampattern formulations. These beampatterns are obtained by solving problem (8) for the radar-only case, and solving (9) and (34) for IRS-assisted ISAC systems with perfect and uncertain channel knowledge cases, respectively, using the proposed FP-SDP-SOCP based algorithm.

Fig. 2 compares the beampattern of radar-only system against the ISAC beampatterns with and without the assistence of IRS. The obtained beampatterns of the no-IRS and the IRS-assisted ISAC scenarios are compared under different required SINR values for the communication users to show the effect of increasing the SINR value. In all cases, the main-beam of the desired beampattern is centered at $0°$ with a 3dB width of $20°$. For the no-IRS scenario, it is obvious that the beampattern exhibits low PSLR, compared to the radar-only beampattern, resulting in a poor radar performance. In contrast, the IRS-assisted ISAC scenario achieves significantly improved beampatterns that are very close to the reference radar-only beampattern. It is also noticeable that the PSLR of the ISAC beampatterns decreases as the SINR level of the communication users increases as more power should be consumed to satisfy the SINR constraints rather than minimizing the Frobenius distance in (9a).

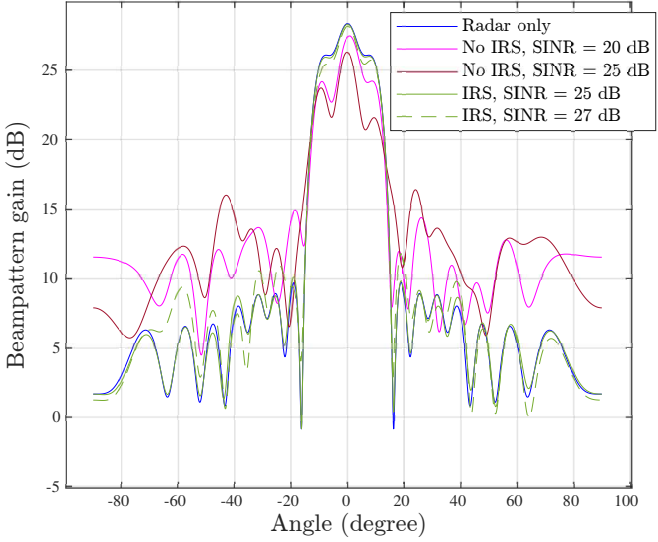

Figure 2: Comparison of the obtained beampatterns under radar-only, ISAC, and IRS-ISAC scenarios with different SINRs

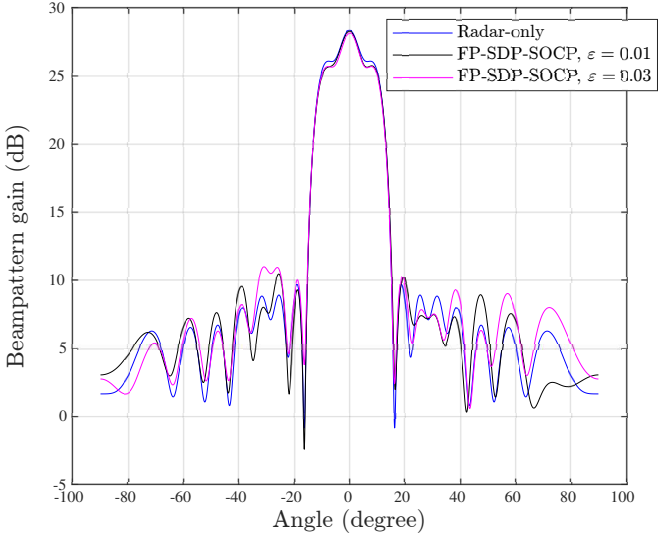

Figure 3: Comparing the obtained beampatterns under robust beamforming with different $\varepsilon$ and SINR = 24dB against the reference radar-only beampattern

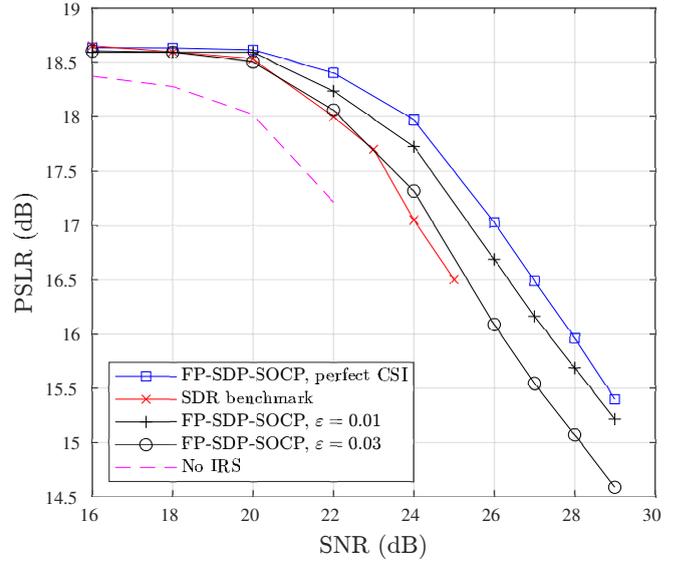

Figure 4: Comparison of the PSLR performance versus the communication SNR for the different ISAC scenarios

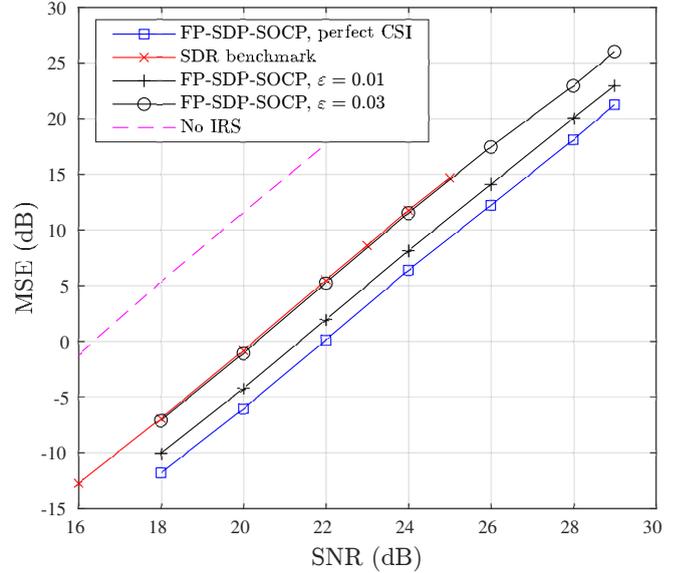

Figure 5: Comparison of the MSE performance versus the communication SNR for the different ISAC scenarios

Fig. 3 compares the the obtained beampatterns for the robust beamforming case with different channel uncertainty radii against the perfect channel knowledge case. Clearly, the figure shows that as the channel uncertainty radius increases, the quality of the beampattern, i.e., PSLR, decreases resulting in performance degradation in the radar service. The PSLR reduction comes at the expense of consuming more power to satisfy the communication SINR constraints in the presence of channel uncertainty. The more the channel uncertainty radius, the more power consumed in satisfying the SINR constraints, and hence the less the beampattern quality.

Fig. 4 and Fig. 5 explore the performance trade-off between radar and communications when using the different beamforming methods for the IRS-assisted ISAC systems. The two figures plot the PSLR and MSE of the generated beampatterns against various SINR values for the communication users. The number of IRS elements in both figures is set as $L = 30$.

Fig. 4 shows the PSLR of the generated radar-communication beampatterns under different scenarios against the SINR value required for the downlink users. It is evident from the figure that as the SINR value increase, the PSLR of the generated beampatterns decrease. The figure also shows a significant performance improvement by using IRS in the ISAC system. Clearly, the IRS-assisted ISAC scenario outperforms the no-IRS scenario by a significant margin, which illustrates the importance of using IRS. Additionally, the graph shows that the proposed FP-SDP-SOCP based beamforming technique outperforms the SDR benchmark in terms of the radar performance. The proposed beamforming technique provides beampatterns with higher PSLR than the benchmark, besides having less complexity. Finally, it is noticeable, as expected, that the proposed robust beamforming technique provides less PSLR than the non-robust one, and the radar performance decrease as the channel uncertainty radius increase.

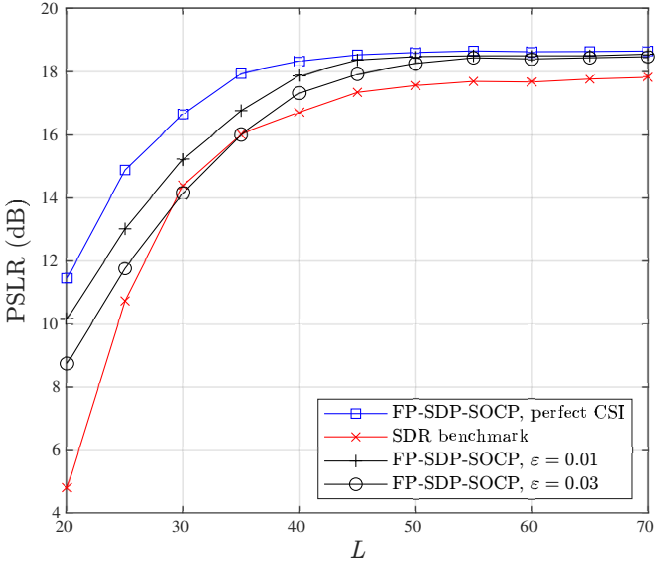

Figure 6: Comparison of the PSLR performance versus the number of IRS reflectors, $L$, for the different ISAC scenarios

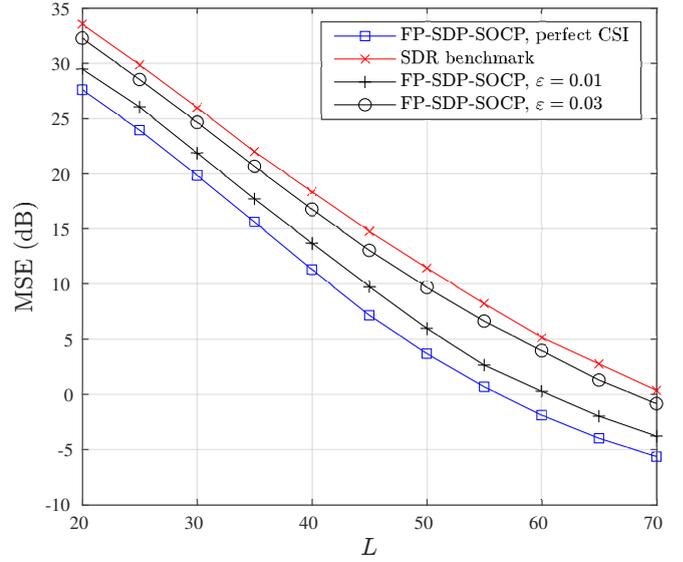

Figure 7: Comparison of the MSE performance versus the number of IRS reflectors, $L$, for the different ISAC scenarios

Fig. 5 shows the MSE of the generated radar-communication beampatterns under different scenarios against the required SINR value for the communication users. Similar trends to those in Fig. 4 are observed in Fig. 5. Specifically, the no-IRS ISAC scenario have the highest MSE compared to the IRS-assisted scenarios. Clearly, the proposed beamforming technique provides lower MSE than the SDR benchmark, besides having lower computational complexity. Also, the MSE of the obtained beampatterns increases in the case of robust beamforming, where the MSE gets higher as the channel uncertainty radius increase. Hence, Fig. 5 reveals a fundamental trade-off between robustness against channel uncertainties and the radar performance.

Fig. 6 shows the PSLR performance trends of the different obtained radar-communication beampatterns versus the number of IRS reflection elements, $L$. The figure shows the effect of increasing the number of IRS reflectors on the radar performance, while fixing the required communication SINR level at 25dB. As the number of IRS reflectors increases, the PSLR of the acquired beampatterns also rises. This improvement can be attributed to the assistance provided by IRS, which enhances the SINR of the communication users. This, in turn, enables the BS to allocate more power to achieve a beampattern that closely matches the desired radar beampattern. Additionally, the figure shows that the proposed low complexity FP-SDP-SOCP based algorithm outperforms the SDR benchmark in terms of PSLR, while having lower computational complexity. Moreover, it is clear that increasing the number of IRS reflectors helps in enhancing the performance of the proposed robust beamforming scheme. However, the radar performance decreases as the channel uncertainty radius increases.

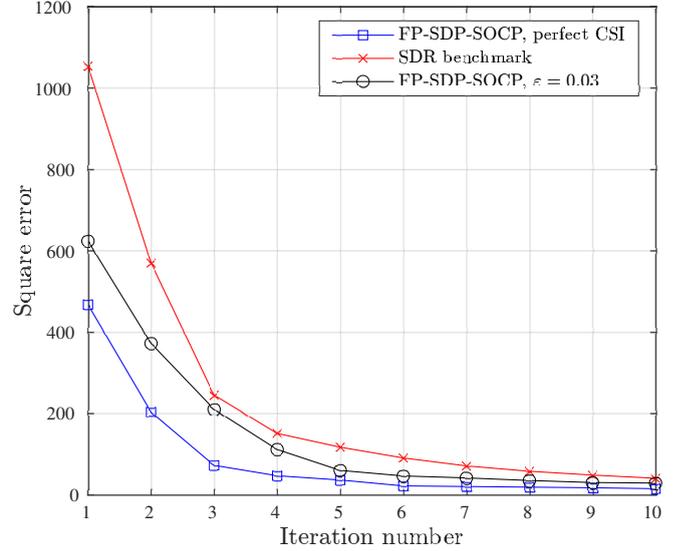

Figure 8: Comparison of the convergence behaviour of the proposed algorithm and its robust version against the SDR benchmark

In Fig. 7, we present the MSE performance of the different obtained beampatterns versus the number of IRS reflectors. The MSE curves in Fig. 7 shows similar trends to those in Fig. 6. It is evident that the MSE performance of the proposed algorithm outperforms the SDR benchmark, and performance of the robust version of the proposed beamforming technique decreases as the channel uncertainty radius increases.

Fig. 8 show the convergence behaviour of the proposed FP-SDP-SOCP algorithm and the robust extension of it. The figure also compare the convergence behaviour of the proposed algorithm against the SDR benchmark. Clearly, the graph shows that the proposed algorithm exhibits faster convergence than the SDR benchmark. Additionally, we proved earlier in the complexity analysis sub-section that each iteration of the proposed algorithm has considerably less complexity compared to the SDR benchmark, which signifies the superiority of the proposed algorithm.

## VII. CONCLUSIONS

In conclusion, we have proposed an efficient and low complexity beamforming technique for IRS-assisted ISAC systems, which can be extended to the robust beamforming case and deal with channel uncertainties. IRS have been proved to enhance the performance of the radar service while maintaining the required communication QoS. By utilizing the proposed algorithm, we could reach better radar beampatterns yielding to better radar performance at a significantly lower complexity. The performance of the IRS-assisted ISAC system is investigated too in the presence of IRS channel uncertainty using the proposed robust beamforming technique. The results reveal that the radar service performance is affected specially at high channel uncertainty radii. However, our proposed robust beamforming algorithm can still provide good radar performance without violating the communication SINR constraints due to the channel uncertainties.